\documentclass{bcp}

\usepackage{amsmath,amssymb,psfig,epsf}

\def\LaTeX{L\kern -.36em\raise .3ex\hbox{\sc a}\kern -.15em T\kern -.1667em%
\lower .7ex\hbox{E}\kern -.125em X}

\begin{document}

\mathclass{ }
\abbrevauthors{P.-H. Chavanis et al.} \abbrevtitle{}

\title{On the analogy between self-gravitating Brownian \\
particles and bacterial populations}

\author{Pierre-Henri Chavanis}
\address{Laboratoire de Physique Th\'eorique, Universit\'e
Paul Sabatier\\ 118 route de Narbonne, 31062 Toulouse Cedex 4,
France\\
E-mail: chavanis@irsamc.ups-tlse.fr}

\author{Magali Ribot}
\address{MAPLY, Universit\'e Claude Bernard\\  B\^atiment 
101, 69622 Villeurbanne Cedex, France\\
E-mail: ribot@maply.univ-lyon1.fr}
\author{Carole Rosier}
\address{LAPCS, Universit\'e Claude Bernard \\ 50 avenue Tony Garnier, 69366 Lyon, France \\
E-mail: rosier@maply.univ-lyon1.fr }

\author{Cl\'ement Sire}
\address{Laboratoire de Physique Th\'eorique, Universit\'e
Paul Sabatier\\ 118 route de Narbonne, 31062 Toulouse Cedex 4,
France\\
E-mail: clement@irsamc.ups-tlse.fr}

\maketitlebcp

\abstract{We develop the analogy between self-gravitating Brownian
particles and bacterial populations. In the high friction limit, the
self-gravitating Brownian gas is described by the Smoluchowski-Poisson
system. These equations can develop a self-similar collapse leading to
a finite time singularity.  Coincidentally, the Smoluchowski-Poisson
system corresponds to a simplified version of the Keller-Segel model
of bacterial populations. In this biological context, it describes the
chemotactic aggregation of the bacterial colonies. We extend these
classical models by introducing a small-scale regularization. In the
gravitational context, we consider a gas of self-gravitating Brownian
fermions and in the biological context we consider finite size
effects. In that case, the collapse stops when the system feels the
influence of the small-scale regularization. A phenomenon of
``explosion'', reverse to the collapse, is also possible.}

\section*{1. Introduction.} Self-gravitating systems such as
globular clusters and elliptical galaxies form a Hamiltonian
system of particles in interaction that can be supposed isolated
in a first approximation \cite{bt}. Since energy is conserved, the
proper statistical description of stellar systems is the microcanonical
ensemble \cite{paddy}. The dynamical evolution of elliptical galaxies is
governed by the Vlasov-Poisson system which corresponds to a
collisionless regime. On the other hand, the kinetic theory of
stars in globular clusters is based on the Landau equation (or the
orbit averaged Fokker-Planck equation) which describes a
collisional evolution. These equations conserve mass and energy.
Furthermore, the Landau equation increases the Boltzmann entropy
(H-theorem) due to stellar encounters. These equations have been
studied for a long time in the astrophysical literature and a
relatively good physical understanding has now been achieved. In
particular, globular clusters can experience core collapse related
to the ``gravothermal catastrophe'' \cite{lbw}.

For systems with long-range interactions, statistical ensembles are
not equivalent. Therefore, it is of conceptual interest to compare the
microcanonical evolution of stellar systems to a canonical model. This
can be achieved by considering a gas of self-gravitating Brownian
particles submitted to a friction with an inert gas and a stochastic
force, in addition to self-gravity \cite{crs}.  This system has a
rigorous canonical structure. In the mean-field approximation, the
self-gravitating Brownian gas model is described by the
Kramers-Poisson system. In a strong friction limit, or for large
times, it reduces to the Smoluchowski-Poisson system. These equations
conserve mass and decrease the Boltzmann free energy. They possess a
rich physical and mathematical structure and can lead to a situation
of ``isothermal collapse'' \cite{iso}, which is the canonical version
of the ``gravothermal catastrophe''.  These equations have not been
considered by astrophysicists because the canonical ensemble is not
the correct description of stellar systems and usual astrophysical
bodies do not experience a friction with a gas (except dust particles
in the solar nebula \cite{planete}).  Yet, it is clear that the
self-gravitating Brownian gas model is of considerable conceptual
interest to understand the strange thermodynamics of systems with
long-range interactions and the inequivalence of statistical
ensembles.

In addition, it turns out that the same type of equations occur in
biology in relation with the chemotactic aggregation of bacterial
populations \cite{murray}. A general model of chemotactic aggregation
has been proposed by Keller \& Segel \cite{keller} in the form of two
coupled differential equations. In some approximation \cite{jager}, this
model reduces to the Smoluchowski-Poisson system, exactly like for
self-gravitating Brownian particles. Therefore, there exists an
isomorphism between self-gravitating Brownian particles and bacterial
colonies. In this paper, we shall develop this analogy in detail. We
shall also propose a modification of the ``standard model'' by
introducing a small-scale regularization. In the gravitational
context, we shall invoke Pauli's exclusion principle and consider a
gas of self-gravitating Brownian fermions. In the biological context,
we shall heuristically account for finite size effects by considering
a lattice model. In that case, the collapse stops when the system
feels the small-scale regularization. An explosion phenomenon, reverse
to the collapse, is also possible. Finally, we shall discuss the
difference between elliptical and parabolic models of bacterial
populations and gravitational systems. We shall also show that 
vortices in two-dimensional turbulence exhibit features similar to
stars and bacteries.

\section*{2. The Hamiltonian $N$-stars problem} 
\begin{figure}
\centerline{
\psfig{figure=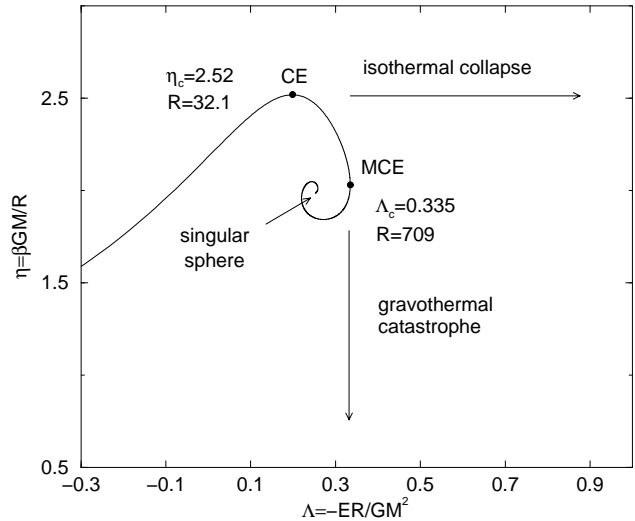,angle=0,height=7cm}}
\caption{Caloric curve for self-gravitating systems exhibiting a ``gravothermal catastrophe'' in the microcanonical ensemble and an ``isothermal collapse'' in the canonical ensemble.}
\label{caloric}
\end{figure}

Consider a 
system of $N$ stars in gravitational interaction. We assume that the
system is isolated so that it conserves mass and energy. The equations
of motion can be cast in a Hamiltonian form
\begin{eqnarray}
{m {d{\bf r}_{i}\over dt}={\partial H\over\partial {\bf v}_{i}},\qquad m {d{\bf v}_{i}\over dt}=-{\partial H\over\partial {\bf r}_{i}}, }
\label{hs1}
\end{eqnarray}
where $H$ is the Hamiltonian
\begin{eqnarray}
H={1\over 2}\sum_{i=1}^{N}mv_{i}^{2}-\sum_{i< j}{Gm^{2}\over |{\bf
r}_{i}-{\bf r}_{j}|}.
 \label{hs2}
\end{eqnarray}
This $N$-body problem is the correct starting point in the description
of globular clusters and elliptical galaxies \cite{bt}.  Since the
system is isolated, the relevant statistical ensemble is the {\it
microcanonical ensemble} \cite{paddy}.  The statistical mechanics of
self-gravitating systems enclosed within a spherical box of radius $R$
was initiated by Antonov \cite{antonov} and Lynden-Bell \& Wood
\cite{lbw}. They found that for $\Lambda\equiv -ER/GM^{2}\ge 0.335$,
corresponding to small energies, there is no maximum entropy state so
that the system must collapse (see Fig. \ref{caloric}). This is the
so-called ``gravothermal catastrophe''.

In the case of elliptical galaxies, encounters between stars are
completely negligible for the timescales of interest and the dynamics
is described by the Vlasov-Poisson system
\begin{equation}
\label{hs3}
{\partial f\over\partial t}+{\bf v}\cdot {\partial f\over\partial {\bf r}}+{\bf F}\cdot {\partial f\over\partial {\bf v}}=0,
\end{equation}
\begin{equation}
\label{hs4}
\Delta\Phi=4\pi G\int f d^{3}{\bf v},
\end{equation}
where $f({\bf r},{\bf v},t)$ is the distribution function, ${\bf
F}({\bf r},t)=-\nabla\Phi$ is the gravitational force and $\Phi({\bf
r},t)$ is the gravitational potential.  By contrast, for globular
clusters, stellar encounters must be taken into account. In that case,
the dynamics is described by the Landau equation
\begin{eqnarray}
\label{hs5}
{\partial f\over\partial t}+{\bf v}\cdot {\partial f\over\partial {\bf r}}+{\bf F}\cdot {\partial f\over\partial {\bf v}}=
{\partial\over\partial v^{\mu}}\int  d^{3}{\bf v}_{1}\ K^{\mu\nu}\biggl\lbrace
f_{1}{\partial f\over\partial v^{\nu}}-f{\partial f_{1}\over\partial v_{1}^{\nu}}\biggr\rbrace,
\end{eqnarray}
where $K^{\mu\nu}$ is the tensor
\begin{equation}
\label{hs6}
K^{\mu\nu}={A\over u}\biggl (\delta^{\mu\nu}-{u^{\mu}u^{\nu}\over u^{2}}\biggr ),
\end{equation}
where $f_{1}=f({\bf r},{\bf v}_{1},t)$, ${\bf u}={\bf v}_{1}-{\bf
v}$ is the relative velocity and $A=2\pi G^{2}m\ln (L_{max}/L_{min})$
is a constant ($L_{max}$ and $L_{min}$ are appropriate lengthscales). This
equation conserves mass $M=\int \rho d^{3}{\bf r}$ and energy $E=\int
f {v^{2}\over 2}d^{3}{\bf r}d^{3}{\bf v}+{1\over 2}\int\rho\Phi
d^{3}{\bf r}$ and satisfies a $H$-theorem ($\dot S\ge 0$) for the
Boltzmann entropy
\begin{equation}
\label{hs7}
S[f]=-\int f\ln f d^{3}{\bf r}d^{3}{\bf v}.
\end{equation}
Therefore, due to the
development of encounters between stars, the system is expected to
relax towards a statistical equilibrium state described by the Maxwell-Boltzmann distribution
\begin{equation}
\label{hs8}
f=A e^{-\beta({v^{2}\over 2}+\Phi)},
\end{equation}
which maximizes the Boltzmann entropy at fixed mass and energy
(microcanonical description).  In fact, this is the case only if the
energy is sufficiently high. Below the Antonov critical energy, there
is no maximum entropy state and the system undergoes {\it core
collapse}. This is a manifestation of the ``gravothermal
catastrophe''. Dynamical models show that the collapse is self-similar
and leads to a finite time singularity (the central density becomes
infinite in a finite time). By solving the orbit averaged
Fokker-Planck equation numerically, Cohn \cite{cohn} finds that the
density profile behaves as $\rho\sim r^{-\alpha}$ with $\alpha=2.23$
at large distances. Alternatively, Lancellotti \& Kiessling
\cite{lance} consider the full Landau-Poisson system and argue that
$\alpha=3$.

\section*{3. The self-gravitating Brownian gas}
The Hamiltonian equations (\ref{hs1})-(\ref{hs2}) describe an isolated
system for which energy is conserved. It can be of interest to study
in parallel a model that is stochastically forced by an external
medium. We thus introduce a system of $N$ Brownian particles in
gravitational interaction described by the Langevin equations
\begin{equation}
\label{bg1}
{d{\bf r}_{i}\over dt}={\bf v}_{i},\quad
{d{\bf v}_{i}\over dt}=-\xi{\bf v}_{i}-\nabla_{i}U({\bf r}_{1},...,{\bf r}_{N})+\sqrt{2D}{\bf R}_{i}(t),
\end{equation}
where $-\xi {\bf v}_{i}$ is a friction force and ${\bf R}_{i}(t)$ is a
white noise satisfying $\langle {\bf R}_{i}(t)\rangle={\bf 0}$ and
$\langle
{R}_{a,i}(t){R}_{b,j}(t')\rangle=\delta_{ij}\delta_{ab}\delta(t-t')$,
where $a,b=1,2,3$ refer to the coordinates of space and $i,j=1,...,N$
to the particles. The particles interact via the gravitational
potential $U({\bf r}_{1},...,{\bf r}_{N})=\sum_{i<j}u({\bf r}_{i}-{\bf
r}_{j})$ where $u({\bf r}_{i}-{\bf r}_{j})=-G/|{\bf r}_{i}-{\bf
r}_{j}|$. We define the inverse temperature $\beta=1/T$ through the
Einstein relation $\xi=D\beta$. The self-gravitating Brownian gas
model has a rigorous canonical structure (see Appendices A and B)
where the temperature $T$ measures the strength of the stochastic
force. The stochastic process (\ref{bg1}) defines a model of gravitational 
dynamics which extends the classical 
Einstein-Smoluchowski Brownian model \cite{risken} to the case 
of stochastic particles {\it in  interaction}. In
this context, the friction is due to the presence of an inert gas and
the stochastic force is due to classical Brownian motion, turbulence
or any other stochastic effect. This model can also be viewed as a
generalization of the Chandrasekhar
\cite{chandra} stochastic model which describes the evolution of a
{\it single} test particle in a stellar cluster at statistical
equilibrium (thermal bath approximation). In that context, the
diffusion and the friction model stellar encounters. More generally,
the friction and the noise can mimick the overall influence of an external
medium (not represented) with which the particles interact.

Starting from the $N$-body Fokker-Planck equation and implementing a
mean-field approximation which is valid in a proper thermodynamic
limit $N\rightarrow +\infty$ with $\eta=\beta GM/R$ fixed, we show in
Appendix B that the distribution function $f({\bf r},{\bf v},t)$
satisfies the Kramers-Poisson system
\begin{equation}
\label{bg2}
{\partial f\over\partial t}+{\bf v}\cdot {\partial f\over\partial {\bf r}}+{\bf F}\cdot {\partial f\over\partial {\bf v}}={\partial \over \partial {\bf v}}\cdot \biggl\lbrace D\biggl\lbrack {\partial f\over\partial {\bf v}}+\beta f{\bf v}\biggr \rbrack \biggr\rbrace,
\end{equation}
\begin{equation}
\label{bg3}
\Delta\Phi=4\pi G\int f d^{3}{\bf v}.
\end{equation}
The Kramers-Poisson system decreases the Boltzmann free energy
\begin{equation}
\label{bg4}
F[f]=E-TS=\int f {v^{2}\over 2}d^{3}{\bf r}d^{3}{\bf v}+{1\over 2}\int\rho\Phi d^{3}{\bf r}+T\int f\ln f d^{3}{\bf r}d^{3}{\bf v}.
\end{equation}
Therefore, the system is expected  to
relax towards a statistical equilibrium state described by the Maxwell-Boltzmann distribution
\begin{equation}
\label{bg5}
f=A e^{-\beta({v^{2}\over 2}+\Phi)},
\end{equation}
which minimizes the Boltzmann free energy at fixed mass and temperature
(canonical description). In fact, this is the case only if the
temperature is sufficiently high. Below a critical temperature
\cite{iso}, the free energy has no minimum and the system undergoes an
``isothermal collapse'' (see Fig. \ref{caloric}). We note that the
equilibrium states of stellar systems (e.g., globular clusters) and
self-gravitating Brownian particles are both given by the isothermal
distribution (\ref{hs8}) or (\ref{bg5}). However, the stability limits
are different in each ensemble because the caloric curve $\beta(E)$
presents turning points (see Fig. \ref{caloric}).  The stability of
stellar systems and self-gravitating Brownian particles differs
in the region of ensemble inequivalence where the specific heat
$C=dE/dT$ is negative
\cite{iso}.

\begin{figure}
\vskip0.6cm
\centerline{
\psfig{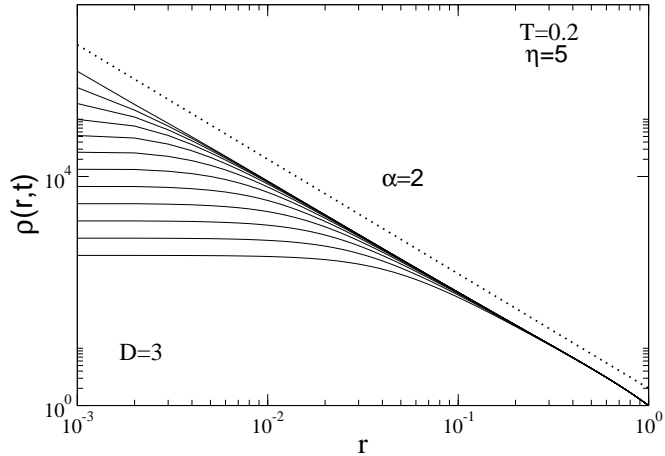}}
\caption{Isothermal collapse in $D=3$. For $\eta=\beta GM/R>2.52$, the system undergoes a finite time singularity leading to a $\rho\sim r^{-2}$ density profile \cite{crs,sire}. A Dirac peak is formed in the post-collapse regime \cite{post}.}
\label{ic3}
\end{figure}

In the high friction limit $\xi\rightarrow +\infty$, or equivalently
for large times $t\gg \xi^{-1}$, we can neglect the inertia of the
particles so that the Langevin equations (\ref{bg1}) take the form
\begin{equation}
\label{bg6}
\xi{d{\bf r}_{i}\over dt}=-\nabla_{i}U({\bf r}_{1},...,{\bf r}_{N})+\sqrt{2D}\ {\bf R}_{i}(t).
\end{equation}
Again using a mean-field approximation (see Appendix B), we can show
that the density $\rho({\bf r},t)$ satisfies the Smoluchowski-Poisson system
\begin{equation}
\label{bg7}
{\partial \rho\over\partial t}=\nabla\cdot\biggl\lbrack {1\over\xi}(T\nabla \rho+\rho\nabla\Phi)\biggr\rbrack,
\end{equation}
\begin{equation}
\label{bg8}
\Delta\Phi=4\pi G\rho.
\end{equation}
The Smoluchowski equation (\ref{bg7}) can also be deduced from the Kramers
equation (\ref{bg2}) by using a method of moments \cite{gt} or a
Chapman-Enskog expansion \cite{lemou}. We can show that it decreases
the Boltzmann free energy
\begin{eqnarray}
{F}[\rho]=T\int \rho \ln\rho\ d^{3}{\bf r}+{1\over
  2}\int\rho\Phi \ d^{3}{\bf r},
\label{bg9}
\end{eqnarray}
obtained from Eq. (\ref{bg4}) by using the fact that the distribution
function is close to the Maxwellian distribution 
\begin{equation}
\label{ntbe}
f=\biggl ({\beta\over 2\pi}\biggr )^{3/2}\rho({\bf r},t) \ e^{-\beta {v^{2}\over 2}},
\end{equation}
in the high friction limit \cite{gt,lemou}.  The equilibrium states
of the Smoluchowski-Poisson system (\ref{bg7})-(\ref{bg8}) are given
by the Boltzmann distribution
\begin{equation}
\label{bg10} \rho=A'e^{-\beta\Phi}.
\end{equation}

\begin{figure}
\vskip0.6cm
\centerline{
\psfig{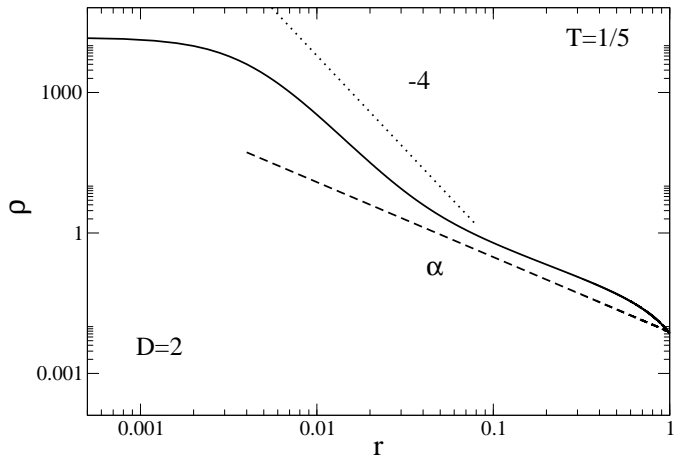}}
\caption{Isothermal collapse in $D=2$. For $\eta=\beta GM>4$, the system develops a Dirac peak surrounded by a halo behaving as $\rho\sim r^{-\alpha}$. For $t\rightarrow t_{coll}$, $\alpha=2$ but the  convergence to this asymptotic value is extremely slow so that an effective exponent $\alpha\simeq 1.3$ is observed for the times achieved in our simulations \cite{sire}.}
\label{ic2}
\end{figure}

In the gravitational context, the Kramers-Poisson system and the
Smoluchowski-Poisson system have been introduced by Wolansky
\cite{wolansky}, Chavanis, Sommeria \& Robert \cite{csr} and Chavanis, Rosier \& Sire \cite{crs} with
different motivations. A physical description of the spherically
symmetric solutions of the Smoluchowski-Poisson system has been given in
\cite{crs,sire,anomalous,post,tcoll} in various dimensions of space
(including the critical dimension $D=2$) and for both pre-collapse and
post-collapse regimes. For sufficiently large mass $M$ or sufficiently
low temperature $T$, the Smoluchowski-Poisson system displays a
self-similar collapse leading to a finite time singularity in $D>2$ (see Fig. \ref{ic3}). The density profile behaves as $\rho\sim
r^{-\alpha}$ with $\alpha=2$ at the collapse time \cite{crs,sire}.
Then, the evolution continues and a Dirac peak (``black hole'') is finally
formed in the post-collapse regime \cite{post}. This is consistent with
predictions of statistical mechanics in the canonical ensemble
\cite{kiessling,iso}. In $D=2$, the evolution is more complex \cite{sire} and
creates a Dirac peak containing a fraction
$T/T_{c}$ of the total mass, surrounded by a $\rho\sim r^{-\alpha}$
halo with an effective scaling exponent converging very slowly to
$\alpha=2$ (see Fig. \ref{ic2}). A review of these results is given in
\cite{bed}.

\section*{4. Self-gravitating Brownian fermions} As discussed previously, 
self-gravitating classical particles have the tendency to develop
finite time singularities. In an attempt to regularize the problem at
high densities, and avoid unphysical infinities, we can invoke quantum
mechanics and use Pauli's exclusion principle. Thus, we shall consider a gas of
self-gravitating Brownian fermions as a regularized model of
gravitational dynamics. The equilibrium states of self-gravitating
fermions and the description of phase transitions in the
self-gravitating Fermi gas (in both microcanonical and canonical
ensembles) have been investigated by Chavanis \cite{pt,ptD}. The
system of self-gravitating Brownian fermions \cite{ribot} can be used
as a simple model to study these phase transitions {\it dynamically}
in the canonical ensemble.

A generalization of the Kramers equation (\ref{bg2}) taking into
account the Pauli exclusion principle is given by \cite{gt}:
\begin{equation}
\label{bf1}
{\partial f\over\partial t}+{\bf v}\cdot {\partial f\over\partial {\bf r}}+{\bf F}\cdot {\partial f\over\partial {\bf v}}={\partial \over \partial {\bf v}}\cdot \biggl\lbrace D\biggl\lbrack {\partial f\over\partial {\bf v}}+\beta f(1-f/\eta_{0}){\bf v}\biggr \rbrack \biggr\rbrace.
\end{equation}
This equation respects the constraint $f\le \eta_{0}\equiv
m^{4}/h^{3}$ ($h$ is Planck constant) at all times and decreases the
Fermi-Dirac free energy
\begin{equation}
\label{bf2}
F=\int f {v^{2}\over 2}d^{3}{\bf r}d^{3}{\bf v}+{1\over 2}\int\rho\Phi d^{3}{\bf r}+T\int \lbrace f\ln f+(\eta_{0}-f)\ln(\eta_{0}-f)\rbrace  d^{3}{\bf r}d^{3}{\bf v}.
\end{equation}
The equilibrium states are given by the Fermi-Dirac distribution function
\begin{equation}
\label{bf3}
f={\eta_{0}\over 1+\lambda e^{\beta ({v^{2}\over 2}+\Phi)}},
\end{equation}
which minimizes the free energy (\ref{bf2}) at fixed mass and temperature.

\begin{figure}
\centerline{
\psfig{figure=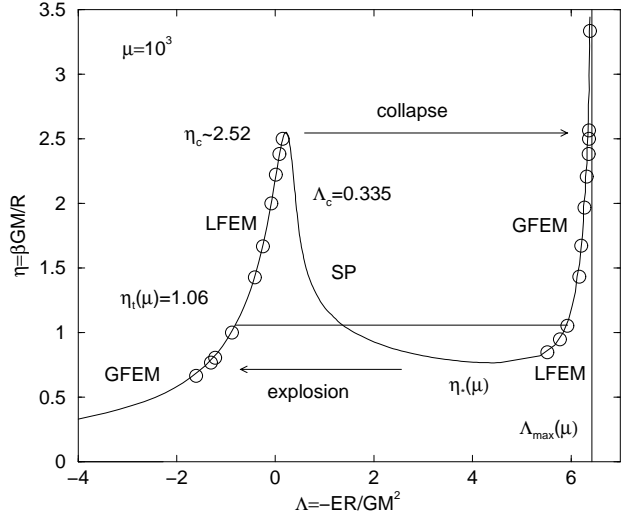,angle=0,height=7cm}}
\caption{Hysteretic cycle in the canonical ensemble for self-gravitating fermions. The system undergoes a ``collapse'' at $T_{c}$ and an ``explosion'' at $T_{*}$. The temperatures of collapse and explosion differ due to the existence of long-lived metastable states (local minima of free energy) \cite{michel}. The circles correspond to the results of numerical simulations of the fermionic Smoluchowski-Poisson system \cite{ribot}.}
\label{hyst}
\end{figure}

Now, considering the high friction limit \cite{gt,lemou}, we can
derive a generalized Smoluchowski equation of the form
\begin{equation}
\label{bf4}
{\partial \rho\over\partial t}=\nabla\cdot\biggl\lbrack {1\over\xi}(\nabla p+\rho\nabla\Phi)\biggr\rbrack,
\end{equation}
where $p(\rho)$ is the local equation of state of the Fermi gas
\begin{equation}
\label{bf5}
\rho={4\pi\sqrt{2}\eta_{0}\over\beta^{3/2}}I_{{1\over 2}}(\lambda'),\qquad p={8\pi\sqrt{2}\eta_{0}\over 3\beta^{{5/2}}}I_{{3\over 2}}(\lambda'),
\end{equation}
and
\begin{equation}
\label{bf6}
I_{n}(t)=\int_{0}^{+\infty}{x^{n}\over 1+te^{x}}dx
\end{equation}
is the Fermi integral. The fermionic Smoluchowski-Poisson system
decreases the free energy
\begin{eqnarray}
{F}[\rho]=\int \rho \int_{0}^{\rho}{p(\rho')\over\rho^{'2}}d\rho'
d^D{\bf r}+{1\over
  2}\int\rho\Phi d^{D}{\bf r},
\label{bf7}
\end{eqnarray}
obtained from Eq. (\ref{bf2}) by using the fact that $f({\bf r},{\bf
v},t)$ is close to the Fermi distribution in the high friction limit
\cite{lemou}. Furthermore, the equilibrium states are obtained by
substituting the relation
\begin{equation}
\label{bf8}
\rho={4\pi\sqrt{2}\eta_{0}\over\beta^{3/2}}I_{{1\over 2}}(\lambda e^{\beta\Phi}),
\end{equation}
in the Poisson equation (\ref{bg8}) and solving for the gravitational potential $\Phi$ \cite{pt}.

\begin{figure}
\centerline{
\psfig{figure=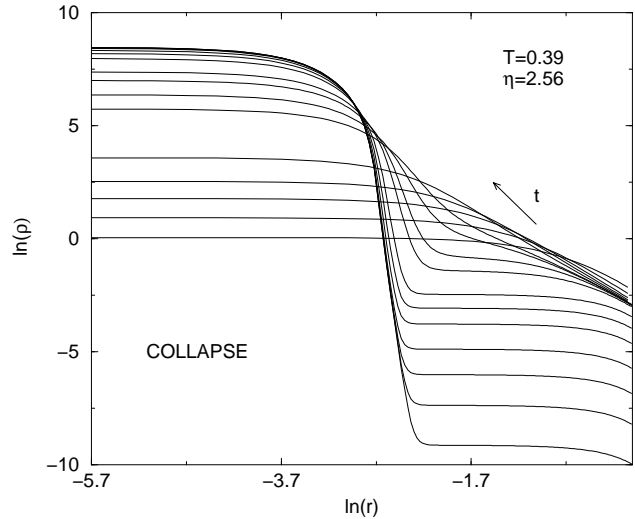,angle=0,height=7cm}}
\caption{Collapse of self-gravitating Brownian fermions for $T<T_{c}$. The system finally forms a ``fermion ball'' (similar to a white dwarf star) surrounded by an isothermal halo (like a vapor). This is a global minimum of free energy in the canonical ensemble.}
\label{collapse}
\end{figure}

\begin{figure}
\centerline{
\psfig{figure=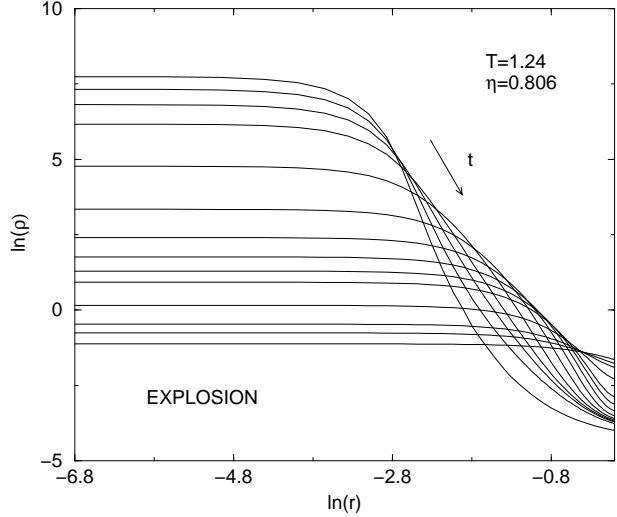,angle=0,height=7cm}}
\caption{Explosion of self-gravitating Brownian fermions for $T>T_{*}$. When diffusion prevails over gravity, the condensed object expands and the system returns to the gaseous phase.  }
\label{explosion}
\end{figure}

The fermionic Smoluchowski-Poisson system has been studied
mathematically by Biler et al. \cite{biler}. Numerical simulations
have been conducted in parallel in \cite{ribot}. In particular, these
simulations reveal an interesting hysteretic cycle (see
Fig. \ref{hyst}) discussed by Chavanis \& Rieutord
\cite{michel}. Below a critical temperature $T_{c}$, coinciding with
the Jeans instability criterion
\cite{crs,iso}, the system collapses under its own gravity 
as in Fig. \ref{ic3}.  However, for self-gravitating fermions, the
collapse stops when the core becomes degenerate (in the sense of the
Fermi-Dirac statistics). The resulting condensed object has the same structure
as a cold white dwarf star (or a ``fermion ball'') in which gravity is
balanced by quantum pressure (see Fig. \ref{collapse}). If now
temperature is increased, the system remains in the condensed phase
until another temperature $T_{*}>T_{c}$ (see Fig. \ref{hyst}) at which
it undergoes an explosion (reverse to the collapse) and returns to the
gaseous phase (see Fig. \ref{explosion}).

\section*{5. Chemotactic aggregation} We shall now point out some analogies between a gas of self-gravitating Brownian particles and the chemotaxis of bacterial populations \cite{murray}. The name chemotaxis refers to
the motion of organisms (amoeba) induced by chemical signals
(acrasin). In some cases, the biological organisms secrete a substance
that has an attractive effect on the organisms themselves. Therefore,
in addition to their diffusive motion, they move systematically along
the gradient of concentration of the chemical they secrete
(chemotactic flux). When attraction prevails over diffusion, the
chemotaxis can trigger a self-accelerating process until a point at
which aggregation takes place. This is the case for the slime mold
{\it Dictyostelium Discoideum} and for the bacteria {\it Escherichia
coli}.

A model of slime mold aggregation has been introduced by Keller \&
Segel \cite{keller} in the form of two PDE's:
\begin{equation}
\label{ca1} {\partial\rho\over\partial t}=\nabla\cdot (D_{2}\nabla\rho)-\nabla\cdot (D_{1}\nabla c),
\end{equation}
\begin{equation}
\label{ca2}{\partial c\over\partial t}=-k(c)c+f(c)\rho+D_{c}\Delta c.
\end{equation}
In these equations $\rho({\bf r},t)$ is the concentration of amoebae and
$c({\bf r},t)$ is the concentration of acrasin. Acrasin is produced
by the amoebae at a rate $f(c)$. It can also be degraded at a rate
$k(c)$. Acrasin diffuse according to Fick's law with a diffusion
coefficient $D_c$. Amoebae concentration changes as a result of an
oriented chemotactic motion in the direction of a positive gradient of
acrasin and a random motion analogous to diffusion. In Eq. (\ref{ca1}),
$D_2(\rho,c)$ is the diffusion coefficient of the amoebae and
$D_1(\rho,c)$ is a measure of the strength of the influence of the
acrasin gradient on the flow of amoebae. This chemotactic drift is the
fundamental process in the problem.

A first simplification of the Keller-Segel model is provided by
the system of equations
\begin{equation}
\label{ca3} {\partial\rho\over\partial t}=D\Delta\rho-\chi\nabla\cdot (\rho\nabla c),
\end{equation}
\begin{equation}
\label{ca4}{\partial c\over\partial t}=D'\Delta c+a \rho-b c,
\end{equation}
where the parameters are positive constants. An additional
simplification, introduced by J\"ager \& Luckhaus \cite{jager},
consists in ignoring the temporal derivative in Eq. (\ref{ca4}). This is
valid in the case where the diffusion coefficient $D'$ is
large. Taking also $b=0$, we obtain
\begin{equation}
\label{ca5} {\partial\rho\over\partial t}=D\Delta\rho-\chi\nabla \cdot (\rho\nabla c),
\end{equation}
\begin{equation}
\label{ca6}\Delta c=-\lambda\rho,
\end{equation}
where $\lambda=a/D'$. Clearly, these equations are isomorphic to the
Smoluchowski-Poisson system (\ref{bg7})-(\ref{bg8}) describing
self-gravitating Brownian particles in a high friction limit. In
particular, the chemotactic flux plays the same role as the
gravitational drift in the overdamped limit of the Brownian model. We
have the correspondance $\Phi\leftrightarrow -{4\pi G\over\lambda}c$,
$\beta\leftrightarrow {\lambda\chi\over 4\pi GD}$ and
$\xi\leftrightarrow{4\pi G\over \lambda\chi}$. Through this analogy,
we can develop an {\it effective} thermodynamical formalism to
investigate the chemotactic problem \cite{gt,bcp}. In particular,
Eq. (\ref{ca5}) is similar to a Fokker-Planck equation and its stationary
solutions are similar to the Boltzmann distribution 
\begin{equation}
\label{gcm1newq} \rho=A e^{{\chi\over D}c},
\end{equation}
which maximizes a Lyapunov functional similar to the Boltzmann free energy.

\section*{6. A generalized chemotactic model} The Keller-Segel
model ignores clumping and sticking effects. However, at the late
stages of the blow-up, when the density of amoebae has reached high
values, finite size effects and stickiness must clearly be taken into
account. As a first step, we propose  to replace the
classical equation (\ref{ca5}) by an equation of the form
\begin{equation}
\label{gcm1} {\partial\rho\over\partial t}=D\Delta\rho-\chi\nabla \cdot (\rho(1-\rho/\sigma_{0})\nabla c),
\end{equation}
which enforces a limitation $\rho\le \sigma_{0}$ on the maximum
concentration of bacteria in physical space \cite{model}. This equation increases
the Lyapunov functional
\begin{equation}
\label{gcm2} {J}[\rho]=-\int \lbrack \rho\ln\rho+(\sigma_{0}-\rho)\ln(\sigma_{0}-\rho)\rbrack d^{3}{\bf r}+{\chi\over 2D}\int \rho c  \ d^{3}{\bf r}.
\end{equation}
In the thermodynamical analogy mentioned above, this functional can be
interpreted as a free energy $J=S-\beta E$ associated with a
Fermi-Dirac entropy in physical space \cite{gt,bcp}.  This form of
entropy can be obtained by introducing a lattice model preventing two
particles to be on the same site. The lattice creates an exclusion
principe in physical space similar to the Pauli exclusion principle in
phase space. Then, $S[\rho]$ can be obtained by a standard
combinatorial analysis respecting this exclusion principle. The
equilibrium states of Eq. (\ref{gcm1}) are given by a Fermi-like
distribution in physical space
\begin{equation}
\label{gcm3} \rho={\sigma_{0}\over 1+\lambda e^{-{\chi\over D}c}},
\end{equation}
which maximizes the effective free energy (\ref{gcm2}) at fixed mass.

It is clear on qualitative grounds that the system of equations
(\ref{gcm1})-(\ref{ca6}) will display exacty the same phenomena as the
system of self-gravitating Brownian fermions (at least in dimension
$D=3$) \cite{model}.  In particular, by tuning the mass of the system,
or more generally the dimensionless parameter $\eta={\lambda\chi
M\over 4\pi DR}$, we can describe an hysteretic cycle similar to the
one depicted in Fig. \ref{hyst}. Above a critical mass, blow-up occurs
until finite size effects come into play and arrest the
collapse. Then, by slowly decreasing the mass of the aggregate, we
reach a critical point $\eta_{*}$ at which an explosion sets in. This
occurs when diffusion prevails over chemotactic drift. This hysteretic
cycle has never been reported in the chemotactic literature and we can
wonder whether it could be observed in biological experiments. In any
case, the constraint $\rho\le\sigma_{0}$ implied by Eq.  (\ref{gcm1})
regularizes the problem and prevents unphysical infinities.

On the other hand, as pointed out by Keller \& Segel \cite{keller},
the diffusion coefficient of amoebae can depend on the density $\rho$,
leading to a situation of anomalous diffusion. For example, the case
where the diffusion coefficient is a power law of the density has been
investigated in \cite{anomalous}. Moreover, the relation between
the concentration of amoebae and acrasin may be more complex that
simply given by the Poisson equation (\ref{ca6}). For example, taking
$b\neq 0$ in the original model, we obtain a relation of the form
\begin{equation}
\label{gcm4}\Delta c-k_{S}^{2}c=-\lambda\rho,
\end{equation}
where $k_{S}^{2}=b/D'$.  The second term is similar to the Debye shielding
in plasma physics. These remarks motivate us to consider a larger
class of drift-diffusion equations of the form
\begin{equation}
\label{gcm5} {\partial \rho\over\partial t}=\nabla\cdot \biggl\lbrace
D\biggl\lbrack \nabla\rho+{\beta\over C''(\rho)} \nabla\int u({\bf r}-{\bf r}')\rho({\bf r}',t)d^{D}{\bf r}'\biggr
\rbrack \biggr\rbrace.
\end{equation}
This generalized class of Fokker-Planck equations has been introduced
by Chavanis \cite{gt,physA,lemou,bcp}. They include an arbitrary diffusion
coefficient $D({\bf r},t)$, an arbitrary convex function $C(\rho)$ and
an arbitrary binary potential of interaction $u({\bf r}-{\bf
r}')$. Equation (\ref{gcm5}) can therefore provide a generalized model
of chemotactic aggregation taking into account anomalous diffusion,
stickening effects and shielding effects of various forms. This
equation increases a Lyapunov functional
\begin{equation}
\label{gcm6} J\lbrack\rho\rbrack =-\int C(\rho)d^{D}{\bf r}-{1\over 2}\beta\int \rho({\bf r},t)u({\bf r}-{\bf r}')\rho({\bf r}',t)d^{D}{\bf r}d^{D}{\bf r}',
\end{equation}
which plays the role of a generalized free energy in an effective
thermodynamical formalism. Finally, the stationary solutions of this
equation are determined by the integrodifferential equation
\begin{equation}
\label{gcm7}C'(\rho)=-\beta\int u({\bf r}-{\bf r}')\rho({\bf r}')d^{D}{\bf r}-\alpha.
\end{equation}
In the limit of short range interactions, the non-local
drift-diffusion equation (\ref{gcm5}) reduces to a form of
Cahn-Hilliard equation
\begin{equation}
\label{gcm8}{\partial\rho\over\partial t}=\nabla\cdot \biggl \lbrace \chi\nabla (\Delta\rho-V'(\rho))\biggr\rbrace,
\end{equation}
where $V(\rho)=-(2/b\beta)C(\rho)-(a/b)\rho^{2}$, $\chi=\beta {b\over
2}{D\over C''(\rho)}$ with $a=\int u(|{\bf x}|) d^{D}{\bf x}$ and $b=
{1\over D} \int u(|{\bf x}|)x^2 d^{D}{\bf x}$ \cite{bcp,lemou}. The
Cahn-Hilliard equation has been extensively studied in the theory of
phase ordering kinetics. In the chemotactic model, the short-range
interaction limit is reached when $k\gg 1$, i.e. in a regime of high
degradation rate. The Cahn-Hilliard equation is known to develop
``domain walls'' and other morphological structures. It would be
interesting to see whether such solutions can be constructed and observed
in the context of bacterial aggregation.

The drift-diffusion equation (\ref{gcm5}) does not take into account
memory effects. However, if we come back to the original Keller-Segel model, the concentration of acrasin $c$ is related to the concentration of amoeba $\rho$ by an equation of the form 
\begin{equation}
\label{gcm9}{\partial c\over\partial t}=D'\Delta c+a \rho-b c,
\end{equation}
which involves a time derivative. Therefore, a more general model of
chemotaxis is represented by the non-Markovian equation
\begin{equation}
\label{gcm10} {\partial \rho\over\partial t}=\nabla\cdot\biggl\lbrace
D\biggl \lbrack \nabla \rho+{\beta\over C''(\rho)}\nabla\int\int_{0}^{t}
u({\bf r}-{\bf r}',t-t')\rho({\bf r}',t')d^{D}{\bf r}'dt'\biggr
\rbrack\biggr\rbrace
\end{equation}
taking into account delay effects \cite{bcp}.

\section*{7. Hyperbolic models} 
\begin{figure}
\centerline{
\psfig{figure=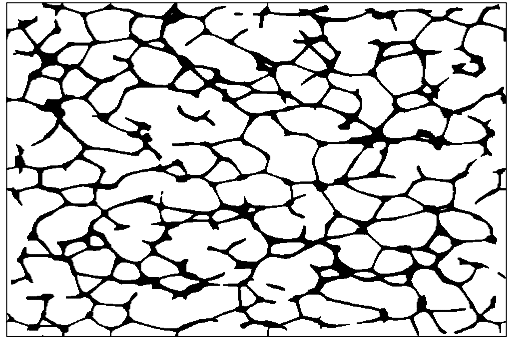,angle=0,height=5cm}}
\caption{Formation of a network in the hyperbolic model (\ref{hm1})-(\ref{hm2}) of chemotactic aggregation \cite{filbet}. We kindly acknowledge F. Filbet for having provided this figure.}
\label{filbet}
\end{figure}

Recent experiments of {\it in vitro} formation of blood vessels show that cells randomly spread on a gel matrix autonomously organize to form a connected vascular network that is interpreted as the beginning of a vasculature. This phenomenon is responsible of angiogenesis, a major actor for the growth of tumors. These networks cannot be explained by the above parabolic models that lead to pointwise blow-up. However, they can be recovered  by certain   hyperbolic models. In particular, Gamba et al. \cite{gamba} describe the evolution of the cell population by the equations
\begin{equation}
\label{hm1} {\partial \rho\over\partial t}+\nabla\cdot (\rho {\bf u})=0,\qquad {\partial {\bf u}\over\partial t}+{\bf u}\cdot\nabla{\bf u}=-\nabla h(\rho)+\mu\nabla c,
\end{equation}
\begin{equation}
\label{hm2}{\partial c\over\partial t}=D'\Delta c+a\rho-bc,
\end{equation}
and show, through extensive numerical simulations, that these
equations develop network patterns that are in good agreement with
experimental results. Additional numerical simulations are performed
in \cite{filbet} (see Fig. \ref{filbet}). 

It can be noted that these equations are similar to the Euler-Jeans equations
\begin{equation}
\label{hm3} {\partial \rho\over\partial t}+\nabla\cdot (\rho {\bf u})=0,\qquad
 {\partial {\bf u}\over\partial t}+{\bf u}\cdot\nabla{\bf u}=-\nabla h(\rho)-\nabla \Phi,
\end{equation}
\begin{equation}
\label{hm4}\Delta\Phi=4\pi G\rho,
\end{equation}
considered in the astrophysical literature (in that context, $h(\rho)$
is the enthalpy).  These equations also describe the emergence of
network-like patterns in the large-scale distribution of masses in the
universe (in that case we need to account for the expansion of the
universe). This is another aspect of the beautiful analogy between
astrophysics and biology.

The connexion between parabolic and hyperbolic models can be made
explicit by introducing a friction force $-\xi {\bf u}$ in the
momentum equation. This friction term can take into account the
influence of the gel matrix on the motion of cells in biology or the
friction with a gas in the case of dust particles moving through the
solar nebula \cite{planete}. In the high friction limit, or for
sufficiently large times $t\gg \xi^{-1}$, the inertia of the particles
can be neglected and the parabolic models (\ref{bg7}) and (\ref{ca5}) 
are recovered.

\section*{8. The Hamiltonian $N$-vortex problem} We finally conclude this
discussion by showing that two-dimensional vortices in fluid mechanics
also share some analogies with self-gravitating systems and bacterial
populations.  Consider a system of $N$ point vortices in 2D
hydrodynamics. The equations of motion can be cast in a Hamiltonian
form
\begin{eqnarray}
\gamma {dx_{i}\over dt}={\partial H\over\partial y_{i}}, \qquad \gamma {dy\over dt}=-{\partial H\over\partial x_{i}},
\label{v1}
\end{eqnarray}
where $H$ is the Hamiltonian
\begin{eqnarray}
H=-{1\over 2\pi}\sum_{i<j}\gamma^{2}\ln |{\bf r}_{i}-{\bf r}_{j}|,
 \label{v2}
\end{eqnarray}
and $\gamma$ the circulation of a point vortex. Since the total energy
is conserved, the correct statistical description is the
microcanonical ensemble.  The statistical mechanics of point vortices
was initiated by Onsager
\cite{onsager}. He showed the existence of negative temperature states
$\beta<0$ at which point vortices cluster in macrovortices. This property is
related to the emergence of large-scale vortices in atmospheric flows
like, e.g., Jupiter's great red spot.

The statistical equilibrium state of a system of point vortices has
been determined by Joyce \& Mongomery \cite{jm} in a meanfield
approximation. It is obtained by maximizing the Boltzmann entropy
\begin{eqnarray}
S=-\int\omega\ln\omega d^{2}{\bf r},
 \label{v3}
\end{eqnarray}
at fixed circulation $\Gamma=\int\omega d^{2}{\bf r}$ and energy $E={1\over 2}\int \omega\psi d^{2}{\bf r}$, where $\omega$ is the vorticity and $\psi$ the streamfunction. This leads to the Boltzmann distribution
\begin{eqnarray}
\omega=Ae^{-\beta\psi},
 \label{v4}
\end{eqnarray} 
where the stream-function $\psi$ is determined self-consistently by substituting the relation (\ref{v4}) in the Poisson equation
\begin{eqnarray}
\omega=-\Delta\psi,
 \label{v5}
\end{eqnarray} 
and solving the resulting Boltzmann-Poisson system. We note that the
equilibrium distribution of point vortices at negative temperatures
(\ref{v4})(\ref{v5}) is similar to the equilibrium distribution of stars
(\ref{hs8}) (\ref{hs4}) and bacteries (\ref{gcm1newq})
(\ref{ca6}). Therefore, there is a deep analogy between
two-dimensional vortices, stellar systems and biological clusters as
first emphasized in \cite{houches,gt}.

A kinetic theory of point vortices has been developed by Chavanis
\cite{kin}, using methods inspired from stellar dynamics and plasma
physics. Starting from the $N$-body Liouville equation, using
projection operator technics and considering axisymmetric flows, he obtained a kinetic equation of the form
\begin{eqnarray}
{\partial \omega\over\partial t}=-{\gamma\over 4 r}{\partial\over\partial
r}\int_{0}^{+\infty}r_{1}dr_{1} \delta(\Omega-\Omega_{1})\ln\biggl \lbrack
1-\biggl ({r_{<}\over r_{>}}\biggr )^{2}\biggr \rbrack\biggl\lbrace {1\over
r}\omega_{1}{\partial \omega\over\partial r}- {1\over r_{1}}\omega{\partial \omega_{1}\over\partial
r_{1}} \biggr\rbrace, \nonumber\\  
\label{v6}
\end{eqnarray}
where $\Omega=\Omega(r,t)$, $\Omega_{1}=\Omega(r_{1},t)$ and $r_{>}$
(resp.  $r_{<}$) is the biggest (resp. smallest) of $r$ and
$r_{1}$. This equation governs the evolution of the average vorticity
$\omega({\bf r},t)=\gamma\rho({\bf r},t)$ which is proportional to the
local density $\rho({\bf r},t)$ of point vortices. The angular velocity
$\Omega(r,t)=\langle V_{\theta}\rangle(r,t)/r$ is related to the
vorticity by
\begin{eqnarray}
\omega ={1\over r}{\partial\over\partial r}(\Omega r^{2}).
\label{v7}
\end{eqnarray}
More general kinetic equations, accounting for memory effects, have
also been derived in \cite{kin} but they are difficult to
study. Equation (\ref{v6}) is the counterpart of the Landau equation
(\ref{hs5}) in gravitational dynamics. The evolution of the vortex
cloud is due to a condition of resonance encapsulated in the
$\delta$-function. This resonance takes into account distant
collisions between point vortices and is manifest only if the profile
of angular velocity is non-monotonic (otherwise the collision term
vanishes). The kinetic equation (\ref{v6}) conserves all the
constraints of the $N$-vortex system (circulation, energy, impulse and
angular momentum) and increases the Boltzmann entropy
(\ref{v3}). However, as discussed in
\cite{kin}, this kinetic equation does {\it not} relax towards the thermal
equilibrium state (\ref{v4}) predicted by the statistical mechanics of
Joyce \& Montgomery \cite{jm}. The reason is not well-understood. This
may be a signature of the lack of ergodicity in the point vortex
gas. This may also correspond to the break-up of the assumptions
leading to Eq. (\ref{v6}). Non-trivial three body correlations may be
necessary to produce further evolution \cite{kin}. However, these
correlations will be manifest on longer timescales and the system can
remain frozen in a sort of metastable state for a very long time.

\section*{9. Brownian vortices} As in the case of self-gravitating 
systems, it is of conceptual interest to compare the microcanonical
evolution of a Hamiltonian system of point vortices to a canonical
model. We thus introduce formally a system of $N$ ``Brownian
vortices'' (the counterpart of self-gravitating Brownian particles)
interacting through a logarithmic potential in two dimensions. By
definition, this system is described by the Langevin equations
\begin{equation}
\label{bv1}
{d{\bf r}_{i}\over dt}=-{\bf z}\times \nabla_{i}U({\bf r}_{1},...,{\bf r}_{N}) -\xi\nabla_{i}U({\bf r}_{1},...,{\bf r}_{N})+\sqrt{2D}{\bf R}_{i}(t),
\end{equation}
where $-\xi\nabla_{i}U({\bf r}_{1},...,{\bf r}_{N})$ is a drift and
${\bf R}_{i}(t)$ is a white noise. The potential of interaction is
given by $U({\bf r}_{1},...,{\bf r}_{N})=\sum_{i<j}u({\bf r}_{i}-{\bf
r}_{j})$ where $u({\bf r}_{i}-{\bf r}_{j})=-{1\over 2\pi}\ln|{\bf
r}_{i}-{\bf r}_{j}|$.  Finally, we define the inverse temperature
$\beta=1/T$ through the Einstein relation $\xi=D\beta$. Since the
temperature $T$ is fixed, the Brownian vortex model has a rigorous
canonical structure. This model can be seen as a generalization of the
stochastic model of Chavanis \cite{drift,kin} which describes the
evolution of a {\it single} test vortex in a bath of field vortices at
statistical equilibrium (this is the counterpart of the Chandrasekhar
\cite{chandra} stochastic model in stellar dynamics). In that case, the drift and the diffusion arise from discrete interactions
with the field vortices which play the role of a thermal
bath. Equation (\ref{bv1}) without the first term can also provide a
stochastic model of chemotaxis.

Starting from the $N$-body Fokker-Planck equation, implementing a
mean-field approximation and following the lines of Appendix B, we
find that the average vorticity $\omega({\bf r},t)=\gamma\rho({\bf
r},t)$ satisfies a Fokker-Planck equation of the form
\begin{equation}
\label{bv2}
{\partial \omega\over\partial t}+{\bf u}\cdot \nabla\omega=\nabla\cdot \lbrack D(\nabla\omega+\beta\gamma\omega\nabla\psi)\rbrack,
\end{equation}
coupled to the Poisson equation
\begin{equation}
\label{bv3}
\Delta\psi=-\omega.
\end{equation}
The drift-diffusion equation (\ref{bv2}) conserves circulation and
increases the free energy $J=S-\beta E$ (canonical
description). Moreover, the Boltzmann distribution (\ref{v4}) is the
only stationary solution of this equation contrary to the kinetic
equation (\ref{v6}) valid for the Hamiltonian system. We note the
analogy with the Smoluchowski-Poisson system (\ref{bg7})-(\ref{bg8})
and (\ref{ca5})-(\ref{ca6}) describing self-gravitating Brownian
particles and bacterial populations. In the case of Brownian point
vortices, the Fokker-Planck equation directly takes the form of a
drift-diffusion equation while for self-gravitating Brownian
particles, the Smoluchowski equation is obtained in a high friction
limit where the motion of the particles is overdamped. We also
emphasize the difference of behaviours between the Brownian vortex
model (\ref{bv1})-(\ref{bv2}) and the Hamiltonian vortex model
(\ref{v1})-(\ref{v6}). In the canonical ensemble in $D=2$, there
exists a critical inverse temperature $\beta_{c}=-{8\pi\over
\gamma\Gamma}$ below which a system of Brownian vortices collapses
\cite{sire}. By contrast, in the microcanononical ensemble,
there is no collapse in $D=2$. This is related to the fact that, in
two dimensions, there exists a maximum entropy state for all values of
energies while the free energy has a minimum only for
$\beta>\beta_{c}$ (the states with $\beta<\beta_c$ are not accessible
in the microcanonical ensemble) \cite{houches}.  Therefore, a naive
canonical description of point vortices in two dimensions does not
describe the true Hamiltonian system correctly. This is a
manifestation of ensemble inequivalence for systems with long-range
interactions, such as point vortices.

\section*{10. Violent relaxation and metaequilibrium states} In the thermodynamical limit $N\rightarrow +\infty$ with $\eta=\beta N\gamma^{2}$
fixed, appropriate to point vortices, the collision term appearing in
the kinetic equation (\ref{v6}) is of order $N^{-1}$. It represents
therefore the first order correction to the Vlasov limit
\cite{houches}. However, in general, the ``collisions'' are negligible
in the point vortex gas and the evolution of the vorticity/density is
described for timescales of interest by the 2D
Euler-Poisson system
\begin{equation}
\label{vr1}
{\partial \omega\over\partial t}+{\bf u}\cdot \nabla\omega=0,
\end{equation}
\begin{equation}
\label{vr2}
\Delta\psi=-\omega.
\end{equation}
This corresponds to the Vlasov regime in stellar dynamics and plasma
physics.  These equations can be derived from the Liouville equation
by assuming that the $N$-body distribution function factorizes as $N$
one-body distribution functions, which is justified in the limit
$N\rightarrow +\infty$ (for fixed $t$) \cite{kin}.  The Euler equation
also describes the inviscid evolution of continuous vorticity flows in
two dimensions (see, e.g.,
\cite{houches}).

\begin{figure}
\centerline{
\psfig{figure=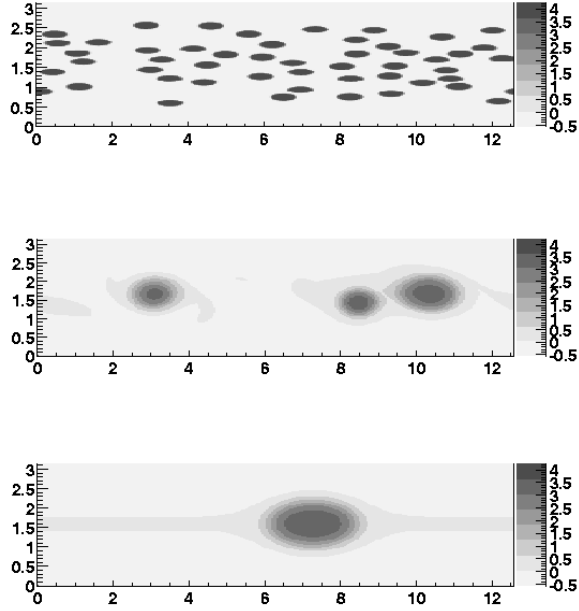,angle=0,height=8cm}}
\caption{Relaxation towards statistical equilibrium in a QG model of Jupiter's great red spot and other jovian vortices (white ovals), using the parameterization (\ref{vr3}) \cite{bd}.  We kindly acknowledge F. Bouchet for having provided this figure.  }
\label{freddy}
\end{figure}

It is known that the 2D Euler-Poisson system develops a complicated
mixing process leading to a metaequilibrium state on the
coarse-grained scale. This process is called violent relaxation. A
small-scale parametrization of the 2D Euler equation can be attempted
by using thermodynamical arguments
\cite{rsmepp} and kinetic theory \cite{quasi}. 
Similar arguments can be developed for the Vlasov-Poisson system
\cite{csr}. In the two-levels approximation of their statistical theory,
Robert \& Sommeria \cite{rsmepp} have proposed a parametrization of
the form
\begin{equation}
\label{vr3}
{\partial \overline{\omega}\over\partial t}+{\bf u}\cdot \nabla\overline{\omega}=\nabla \cdot \lbrack D(\nabla\overline{\omega}+\beta(t)(\overline{\omega}-\sigma_{-1})(\sigma_{1}-\overline{\omega})\nabla\psi)\rbrack,
\end{equation}
where $\overline{\omega}$ and $\psi$ are coupled through the Poisson
equation (\ref{vr2}) and $\beta(t)$ evolves with time so as to satisfy
the conservation of energy $\dot E=0$. This equation respects the
constraints $\sigma_{-1}\le \overline{\omega}\le
\sigma_{1}$ imposed by the Euler equation on the coarse-grained vorticity and converges towards the equilibrium distribution 
\begin{equation}
\label{vr3eq}
\overline{\omega}=\sigma_{-1}+{\sigma_{1}-\sigma_{-1}\over 1+\lambda e^{\beta(\sigma_{1}-\sigma_{-1})\psi}},
\end{equation}
which maximizes the mixing entropy
\begin{equation}
\label{vr3mix}
S=-\int \lbrack p\ln p+(1-p)\ln (1-p)\rbrack d^{2}{\bf r},
\end{equation}
with $\overline{\omega}=p\sigma_{1}+(1-p)\sigma_{-1}$, at fixed
circulation and energy. An example of relaxation towards
statistical equilibrium is shown in
Fig. \ref{freddy} in relation with the formation of large-scale
vortices in the jovian atmosphere \cite{bd}. We note the analogy between
Eq. (\ref{vr3}) and the model (\ref{gcm1}) proposed for the
chemotactic aggregration of bacterial populations (consider in
particular the case $\sigma_{-1}=0$). One important difference is that $\beta$
in Eq. (\ref{vr3}) is not fixed but evolves with time so as to satisfy
the conservation of energy. Hence, Eq. (\ref{vr3}) describes a
microcanonical situation while Eq. (\ref{gcm1}) can be associated with
a canonical situation \cite{gt,bcp}. The well-posedness of Eq. (\ref{vr3}) has been established in \cite{crlr1,crlr2}.

In the quasi-geostrophic approximation appropriate to geophysical
fluid dynamics \cite{pedlosky}, Eqs. (\ref{vr1}) and (\ref{vr3})
remain valid provided that the vorticity $\omega$ is replaced by the
potential vorticity ${q}$ which is related to $\psi$ by a relation of
the form
\begin{equation}
\label{vr4}
q=-\Delta\psi+{1\over R^{2}}\psi+h(y).
\end{equation}
In this equation, $h(y)$ represents a topography and $R$ is the
so-called Rossby radius. It takes into account the deformation of the
surface of the fluid. This equation should be compared with the
similar equation (\ref{gcm4}) in the chemotactic problem. In the limit
of small Rossby radius $R\rightarrow 0$, the relaxation equation
(\ref{vr3}) becomes equivalent to the Cahn-Hilliard equation as
discussed in Ref.
\cite{pa}. In that case, its stationary solutions describe ``domain
walls''. These solutions precisely account  for the jet structure of
Jupiter's great red spot \cite{bs}.

The parametrization of Robert \& Sommeria \cite{rsmepp} is complicated
to implement in the case of realistic initial conditions with a large
number of vorticity levels. In order to go beyond the two-levels
approximation while leaving the problem tractable, Chavanis \cite{gt,pa} has
proposed an equation of the form
\begin{equation}
\label{vr5}
{\partial {q}\over\partial t}+{\bf u}\cdot \nabla {q}=\nabla \cdot \biggl \lbrace D\biggl\lbrack \nabla {q}+{\beta(t)\over C''(q)}\nabla\psi\biggr\rbrack \biggr\rbrace,
\end{equation}
where $C$ is a convex function. The relation between the potential vorticity and the stream-function is written in the general form
\begin{equation}
\label{vr6}
\psi({\bf r},t)=\int u({\bf r}-{\bf r}') q({\bf r}',t)d^{2}{\bf r}'.
\end{equation}
The relaxation equation (\ref{vr5}) provides a simplified
parametrization of the 2D Euler-Poisson system, where the function
$C(q)$ has to be adapted to the physical context (ocean dynamics,
jovian atmosphere,...). Alternatively, it can be used as a numerical
algorithm to construct arbitrary nonlinearly dynamically stable
stationary solutions of the 2D Euler equation, specified by a
relationship $C'(q)=-\beta\psi-\alpha$, in an attempt to reproduce
observed phenomena
\cite{gt,pa}.

\section*{11. Conclusion} In this paper, we have emphasized the analogy between self gravitating
systems, two-dimensional vortices and bacterial populations. This
completes the discussion initiated in \cite{houches,gt}. We have
introduced or quoted different types of models and kinetic
equations. Some models have a rigorous physical justification. Others
have a more academic or conceptual interest.  The common point between
these models is the long-range nature of the interactions leading to
non-local kinetic equations or non-local Fokker-Planck equations
\cite{gt,physA,bcp}. They correspond either to a microcanonical
situation (fixed energy) or a canonical one (fixed temperature).  These equations have a very rich mathematical and
physical structure associated with phase transitions, blow-up
phenomena and morphogenesis. It is clear that these equations have a
``fundamental'' character and that they deserve to be studied in
detail. This systematic study has been undertaken by the
authors in \cite{crs,sire,anomalous,post,tcoll,ribot}.

\vskip0.5cm
{\bf Acknowledgements} One of us (P.-H. C) would like to thank P. Lauren\c
cot for interesting discussions concerning hyperbolic models of chemotaxis.

\section*{Appendix A: The canonical statistical equilibrium state }\

\vskip4pt plus2pt

In this Appendix, we use the equilibrium BBGKY hierarchy and a mean-field
approximation to determine the statistical equilibrium state of
particles in interaction in the canonical ensemble (a similar approach can be developped in the microcanonical ensemble \cite{bbgky}). The canonical
distribution in physical space is given by
\begin{equation}
\label{bgk1} P_{N}({\bf r}_{1},...,{\bf r}_{N})={1\over
Z(\beta)}e^{-\beta U({\bf r}_{1},...,{\bf r}_{N})},
\end{equation}
where $Z(\beta)$ is the partition function
\begin{equation}
\label{bgk2}  Z(\beta)=\int e^{-\beta U({\bf r}_{1},...,{\bf
r}_{N})}d^D{\bf r}_1...d^D{\bf r}_N.
\end{equation}
In practice, we are not interested by the $N$-body distribution
function $P_{N}({\bf r}_{1},...,{\bf r}_{N})$ but rather by the
reduced densities such as
\begin{equation}
\label{bgk3}P_{1}({\bf r})=\int P_{N}({\bf r},{\bf r}_2,...,{\bf
r}_{N})d^D{\bf r}_2...d^D{\bf r}_N,
\end{equation}
\begin{equation}
\label{bgk4}P_{2}({\bf r},{\bf r}')=\int P_{N}({\bf r},{\bf
r}',{\bf r}_3,...,{\bf r}_{N})d^D{\bf r}_3...d^D{\bf r}_N.
\end{equation}
Assuming that the particles are identical, the average density in ${\bf r}$ is 
\begin{equation}
\label{a3}
\rho({\bf r})=\biggl \langle\sum_{i}\delta ({\bf r}_{i}-{\bf r})\biggr \rangle=N P_{1}({\bf r}).
\end{equation}
Taking the derivative of Eq. (\ref{bgk1}) with respect to ${\bf r}_{1}$, we get
\begin{equation}
\label{bgk5}
{\partial P_{N}\over\partial {\bf r}_{1}}({\bf r}_{1},...,{\bf r}_{N})=-\beta P_{N}{\partial U\over\partial {\bf r}_{1}}({\bf r}_{1},...,{\bf r}_{N}).
\end{equation}
We now integrate over ${\bf r}_{2},...,{\bf r}_{N}$ to obtain
\begin{equation}
\label{bgk6}
{\partial P_{1}\over\partial {\bf r}_{1}}({\bf r}_{1})=-\beta \int\prod_{i=2}^{N} d^{D}{\bf r}_{i} \ P_{N}({\bf r}_{1},...,{\bf r}_{N}){\partial U\over\partial {\bf r}_{1}}({\bf r}_{1},...,{\bf r}_{N}).
\end{equation}
For a binary potential of interaction
\begin{equation}
\label{bgk7} U({\bf r}_{1},...,{\bf r}_{N})=\sum_{i<j}u({\bf
r}_{i}-{\bf r}_{j}), \qquad u({\bf r})=u(-{\bf r}),
\end{equation}
the foregoing expression can be rewritten
\begin{equation}
\label{bgk8}
{\partial P_{1}\over\partial {\bf r}_{1}}({\bf r}_{1})=-\beta\sum_{j=2}^{N} \int\prod_{i=2}^{N} d^{D}{\bf r}_{i} \ P_{N}({\bf r}_{1},...,{\bf r}_{N}){\partial u\over\partial {\bf r}_{1}}({\bf r}_{1}-{\bf r}_{j}).
\end{equation}
Since the particles are identical
\begin{equation}
\label{bgk9}
{\partial P_{1}\over\partial {\bf r}_{1}}({\bf r}_{1})=-\beta (N-1) \int\prod_{i=2}^{N} d^{D}{\bf r}_{i} \ P_{N}({\bf r}_{1},...,{\bf r}_{N}){\partial u\over\partial {\bf r}_{1}}({\bf r}_{1}-{\bf r}_{2}),
\end{equation}
or, equivalently,
\begin{equation}
\label{bgk10}
{\partial P_{1}\over\partial {\bf r}_{1}}({\bf r}_{1})=-\beta (N-1) \int d^{D}{\bf r}_{2} \ P_{2}({\bf r}_{1},{\bf r}_{2}){\partial u\over\partial {\bf r}_{1}}({\bf r}_{1}-{\bf r}_{2}).
\end{equation}
For $N\gg 1$ such that $\beta N u\sim 1$, which defines the proper
thermodynamic limit of this nonextensive system, it can be shown
\cite{bbgky} that the mean field approximation is exact, i.e.
\begin{equation}
\label{bgk11}
P_{2}({\bf r}_{1},{\bf r}_{2})=P_{1}({\bf r}_{1})P_{1}({\bf r}_{2})+O(1/N).
\end{equation}
Thus, Eq. (\ref{bgk10}) becomes
\begin{equation}
\label{bgk12}
{\partial P_{1}\over\partial {\bf r}_{1}}({\bf r}_{1})=-\beta N P_{1}({\bf r}_{1}) {\partial\over\partial {\bf r}_{1}}\int d^{D}{\bf r}_{2} \ P_{1}({\bf r}_{2})u({\bf r}_{1}-{\bf r}_{2}).
\end{equation}
Introducing  the mean-field potential 
\begin{equation}
\label{bgk13}
\Phi({\bf r})=\int d^{D}{\bf r}' \ \rho({\bf r}')u({\bf r}-{\bf r}'),
\end{equation}
produced by the smooth density $\rho({\bf r})=NP_{1}({\bf r})$, we obtain
\begin{equation}
\label{bgk14}
{\partial P_{1}\over\partial {\bf r}}({\bf r})=-\beta P_{1}({\bf r}) {\partial\Phi\over\partial {\bf r}}({\bf r}).
\end{equation}
After integration, we find that the density of particles is given by the Boltzmann distribution
\begin{equation}
\label{bgk15} \rho({\bf r})=A e^{-\beta \Phi({\bf r})},
\end{equation}
where $\Phi$ is related to $\rho$ via Eq. (\ref{bgk13}). Hence, Eq. (\ref{bgk15}) must be regarded as an integro-differential equation.

We now consider the canonical distribution in phase space
\begin{equation}
\label{bgk16} P_{N}({\bf r}_{1},{\bf v}_{1},...,{\bf r}_{N},{\bf
v}_{N})={1\over Z(\beta)}e^{-\beta \bigl\lbrace
\sum_{i=1}^{N}{v_{i}^{2}\over 2}+U({\bf r}_{1},...,{\bf
r}_{N})\bigr\rbrace}.
\end{equation}
Noting that the  velocity distribution is Gaussian, we find that Eq. (\ref{bgk15}) is now replaced by the Maxwell-Boltzmann distribution
\begin{equation}
\label{bgk21} f({\bf r},{\bf v})=A e^{-\beta ({v^{2}\over
2}+\Phi({\bf r}))},
\end{equation}
where, again, $\Phi$ is related to $\rho$ via Eq. (\ref{bgk13}). The
distribution function (\ref{bgk21}) can also be obtained by minimizing
the Boltzmann free energy $F_B[f]$ at fixed mass $M$ and temperature
$T$ (see Sec. 3). This method provides a condition of thermodynamical
{stability} $\delta^{2}F\ge 0$, which is not captured by the
equilibrium BBGKY hierarchy. Indeed, the thermodynamical stability is
related to the dynamical stability with respect to the $N$-body
Fokker-Planck equation (see Appendix B) corresponding to the time
dependent BBGKY hierarchy (see also \cite{gt}).

\section*{Appendix B: Brownian particles in interaction}\

\vskip4pt plus2pt

{\bf B1.} {\it Stochastic processes in physical space} In this
Appendix, we derive the mean-field Fokker-Planck equation (\ref{bg7})
describing a gas of Brownian particles in interaction, using the
method of Martzel \& Aslangul \cite{martzel,mthese}. First, we
consider a system of Brownian particles in physical space. The state
of the system is completely specified by the $N$-body distribution
function $P_{N}({\bf r}_{1},...,{\bf r}_{N},t)$ which gives the
density probability of finding at time $t$ the first particle in ${\bf
r}_{1}$, the second in ${\bf r}_{2}$ etc. We assume that the evolution
of the system is governed by the stochastic process
\begin{equation}
\label{a4}
{d{\bf r}_{i}\over dt}=-\mu\nabla_{i}U({\bf r}_{1},...,{\bf r}_{N})+\sqrt{2D_{*}}{\bf R}_{i}(t),
\end{equation}
where ${\bf R}_{i}(t)$ is a
white noise satisfying $\langle {\bf R}_{i}(t)\rangle={\bf 0}$ and
$\langle
{R}_{a,i}(t){R}_{b,j}(t')\rangle=\delta_{ij}\delta_{ab}\delta(t-t')$,
where $a,b=1,...,D$ refer to the coordinates of space and
$i,j=1,...,N$ to the particles. The particles interact via the
potential $U({\bf r}_{1},...,{\bf r}_{N})=\sum_{i<j}u({\bf r}_{i}-{\bf
r}_{j})$.

For a general Markovian process, we have
\begin{eqnarray}
\label{a5}
P_{N}({\bf r}_{1},...,{\bf r}_{N},t+\Delta t)=\int d^{D}(\Delta {\bf r}_{1})...d^{D}(\Delta {\bf r}_{N})\ P_{N}({\bf
r}_{1}-\Delta {\bf r}_{1},...,{\bf r}_{N}-\Delta {\bf
r}_{N},t)\nonumber\\
\times w({\bf r}_{1}-\Delta {\bf r}_{1},...,{\bf
r}_{N}-\Delta {\bf r}_{N}|\Delta {\bf r}_{1},...,\Delta {\bf r}_{N}),
\end{eqnarray}
where $w$ is the transition probability. Expanding the right hand side in Taylor series and introducing the Kramers-Moyal moments
\begin{eqnarray}
\label{a6}
M_{n_{1}...n_{N}}({\bf r}_{1},...,{\bf r}_{N})={\rm lim}_{\Delta
t\rightarrow 0}{1\over \Delta t\ n_{1}!...n_{N}!}\int d^{D}(\Delta {\bf
r}_{1})...d^{D}(\Delta {\bf r}_{N}) \nonumber\\ 
\times (-\Delta {\bf
r}_{1})^{n_{1}}...(-\Delta {\bf r}_{N})^{n_{N}} w({\bf r}_{1},...,{\bf r}_{N}|\Delta {\bf r}_{1},...,\Delta {\bf
r}_{N}),
\end{eqnarray} 
we get the N-body Fokker-Planck equation
\begin{equation}
\label{a7}
{\partial P_{N}\over\partial t}=\sum_{n_{1}...n_{N}}{\partial^{n_{1}}\over\partial {\bf r}_{1}^{n_{1}}}...{\partial^{n_{N}}\over\partial {\bf r}_{N}^{n_{N}}}\lbrace M_{n_{1}...n_{N}}P_{N}\rbrace,
\end{equation}
where the sum runs over all indices such that $\sum_{i}n_{i}\ge 1$. For the stochastic process (\ref{a4}), only a few moments do not vanish, namely
\begin{equation}
\label{a8}
M_{0...n_{i}=1...0}=\mu\nabla_{i}U({\bf r}_{1},...,{\bf r}_{N}),
\end{equation}
\begin{equation}
\label{a9}
M_{0...n_{i}=2...0}=D_{*}.
\end{equation}
Substituting these results in the $N$-body Fokker-Planck equation (\ref{a7}), we obtain
\begin{equation}
\label{a10}
{\partial P_{N}\over\partial t}=\sum_{i=1}^{N} {\partial\over\partial {\bf r}_{i}}\biggl\lbrack  D_{*}{\partial P_{N}\over\partial {\bf r}_{i}}+\mu P_{N}{\partial\over\partial {\bf r}_{i}}U({\bf r}_{1},...,{\bf r}_{N})\biggr\rbrack.
\end{equation}
We note that the stationary solutions of this equation correspond to
the canonical equilibrium distribution (\ref{bgk1}) provided that
$\mu=D_{*}\beta$ (Einstein relation). Therefore, a gas of Brownian
particles in interaction has a rigorous canonical
structure. Integrating over ${\bf r}_{2}$,...,${\bf r}_{N}$, we obtain
\begin{eqnarray}
\label{a11}
{\partial P_{1}\over\partial t}= {\partial\over\partial {\bf r}_{1}}\biggl\lbrack  D_{*} {\partial P_{1}\over\partial {\bf r}_{1}} +\mu (N-1)\int d^{D}{\bf r}_{2} P_{2}({\bf r}_{1},{\bf r}_{2},t){\partial\over\partial {\bf r}_{1}}u({\bf r}_{1}-{\bf r}_{2})\biggr\rbrack.
\end{eqnarray}
Now, implementing a mean-field approximation which is valid in the
proper thermodynamic limit $N\gg 1$ with $\beta N u\sim 1$ \cite{bbgky}:
\begin{equation}
\label{a12}
P_{2}({\bf r}_{1},{\bf r}_{2},t)=P_{1}({\bf r}_{1},t)P_{1}({\bf r}_{2},t)+O(1/N),
\end{equation}
the foregoing equation can be rewritten
\begin{equation}
\label{a13}
{\partial P_{1}\over\partial t}= {\partial\over\partial {\bf r}_{1}}\biggl\lbrack D_{*} {\partial P_{1}\over\partial {\bf r}_{1}}+\mu P_{1} {\partial\Phi\over\partial {\bf r}_{1}}\biggr\rbrack,
\end{equation}
which is clearly the same as
\begin{equation}
\label{a14}
{\partial \rho\over\partial t}= \nabla\lbrack D_{*} (\nabla\rho+\beta \rho \nabla\Phi)\rbrack,
\end{equation}
where $\Phi({\bf r},t)$ is related to $\rho({\bf r},t)$ as in Eq. (\ref{bgk13}).

\vskip4pt plus2pt

{\bf B2.} {\it Stochastic processes in phase space} We now consider
the generalization of the preceding problem to the case of $N$-coupled
Brownian equations in phase space described by the stochastic process
\begin{eqnarray}
\label{b1}
{d{\bf r}_{i}\over dt}={\bf v}_{i},\quad
{d{\bf v}_{i}\over dt}=-\xi{\bf v}_{i}-\nabla_{i}U({\bf r}_{1},...,{\bf r}_{N})+\sqrt{2D}{\bf R}_{i}(t),
\end{eqnarray}
where  ${\bf R}_{i}(t)$ is a white noise. We start from the Markov process, 
\begin{eqnarray}
\label{b2}
P_{N}({\bf r}_{1},{\bf v}_{1},...,{\bf r}_{N},{\bf v}_{N},t+\Delta t)=\int d^{D}(\Delta {\bf r}_{1})d^{D}(\Delta {\bf v}_{1})...d^{D}(\Delta {\bf r}_{N})d^{D}(\Delta {\bf v}_{N})\nonumber\\
\times P_{N}({\bf
r}_{1}-\Delta {\bf r}_{1},{\bf v}_{1}-\Delta {\bf v}_{1},...,{\bf r}_{N}
-\Delta {\bf r}_{N},{\bf v}_{N}-\Delta {\bf v}_{N},t)\nonumber\\
\times w({\bf r}_{1}-\Delta {\bf r}_{1},{\bf v}_{1}-\Delta {\bf v}_{1},...,{\bf
r}_{N}-\Delta {\bf r}_{N},{\bf
v}_{N}-\Delta {\bf v}_{N} |\Delta {\bf r}_{1},\Delta {\bf v}_{1},...,\Delta {\bf r}_{N},\Delta {\bf v}_{N}    ).\nonumber\\
\end{eqnarray}
Now, the transition probability can be written
\begin{eqnarray}
\label{b3}
 w({\bf r}_{1},{\bf v}_{1},...,{\bf
r}_{N},{\bf
v}_{N} |\Delta {\bf r}_{1},\Delta {\bf v}_{1},...,\Delta {\bf r}_{N},\Delta {\bf v}_{N})=\nonumber\\
\delta(\Delta {\bf r}_{1}-{\bf v}_{1}\Delta t)...\delta(\Delta {\bf r}_{N}-{\bf v}_{N}\Delta t)\psi({\bf r}_{1},{\bf v}_{1},...,{\bf
r}_{N},{\bf
v}_{N} |\Delta {\bf v}_{1},...,\Delta {\bf v}_{N}).
\end{eqnarray}
The integration over $\Delta{\bf r}_{1}$,...,$\Delta{\bf r}_{N}$ is straightforward and yields
\begin{eqnarray}
\label{b4}
P_{N}({\bf r}_{1},{\bf v}_{1},...,{\bf r}_{N},{\bf v}_{N},t+\Delta t)=\int d^{D}(\Delta {\bf v}_{1})...d^{D}(\Delta {\bf v}_{N})\nonumber\\
\times P_{N}({\bf
r}_{1}-{\bf v}_{1}\Delta t,{\bf v}_{1}-\Delta {\bf v}_{1},...,{\bf r}_{N}
-{\bf v}_{N}\Delta t,{\bf v}_{N}-\Delta {\bf v}_{N},t)\nonumber\\
\times \psi({\bf r}_{1}-{\bf v}_{1}\Delta t,{\bf v}_{1}-\Delta {\bf v}_{1},...,{\bf
r}_{N}-{\bf v}_{N}\Delta t,{\bf
v}_{N}-\Delta {\bf v}_{N} |\Delta {\bf v}_{1},...,\Delta {\bf v}_{N}    ),
\end{eqnarray}
which is equivalent to
\begin{eqnarray}
\label{b5}
P_{N}({\bf r}_{1}+{\bf v}_{1}\Delta t,{\bf v}_{1},...,{\bf r}_{N}+{\bf v}_{N}\Delta t,{\bf v}_{N},t+\Delta t)=\int d^{D}(\Delta {\bf v}_{1})...d^{D}(\Delta {\bf v}_{N})\nonumber\\
\times P_{N}({\bf
r}_{1},{\bf v}_{1}-\Delta {\bf v}_{1},...,{\bf r}_{N},{\bf v}_{N}-\Delta {\bf v}_{N},t)\nonumber\\
\times \psi({\bf r}_{1},{\bf v}_{1}-\Delta {\bf v}_{1},...,{\bf
r}_{N},{\bf
v}_{N}-\Delta {\bf v}_{N} |\Delta {\bf v}_{1},...,\Delta {\bf v}_{N}    ).
\end{eqnarray}
Expanding the right hand side in Taylor series and introducing the Kramers-Moyal moments
\begin{eqnarray}
\label{b6}
M_{n_{1}...n_{N}}({\bf r}_{1},{\bf v}_{1},...,{\bf r}_{N},{\bf v}_{N})={\rm lim}_{\Delta
t\rightarrow 0}{1\over \Delta t\ n_{1}!...n_{N}!}\int d^{D}(\Delta {\bf
v}_{1})...d^{D}(\Delta {\bf v}_{N})\nonumber\\
 \times (-\Delta {\bf
v}_{1})^{n_{1}}...(-\Delta {\bf v}_{N})^{n_{N}} 
 \psi({\bf r}_{1},{\bf v}_{1},...,{\bf r}_{N},{\bf v}_{N}|\Delta {\bf v}_{1},...,\Delta {\bf
v}_{N}),
\end{eqnarray} 
we obtain the N-body Fokker-Planck equation
\begin{equation}
\label{b7}
{\partial P_{N}\over\partial t}+\sum_{i=1}^{N}{\bf v}_{i}{\partial P_{N}\over\partial {\bf r}_{i}}=\sum_{n_{1}...n_{N}}{\partial^{n_{1}}\over\partial {\bf v}_{1}^{n_{1}}}...{\partial^{n_{N}}\over\partial {\bf v}_{N}^{n_{N}}}\lbrace M_{n_{1}...n_{N}}P_{N}\rbrace,
\end{equation}
where the sum runs over all indices such that $\sum_{i}n_{i}\ge 1$. For the stochastic process (\ref{b1}), only a few moments do not vanish, namely
\begin{equation}
\label{b8}
M_{0...n_{i}=1...0}=\xi {\bf v}_{i}+\nabla_{i}U({\bf r}_{1},...,{\bf r}_{N}),
\end{equation}
\begin{equation}
\label{b9}
M_{0...n_{i}=2...0}=D.
\end{equation}
Substituting these results in the $N$-body Fokker-Planck equation (\ref{b7}), we get
\begin{eqnarray}
\label{b10}
{\partial P_{N}\over\partial t}+\sum_{i=1}^{N}\biggl ({\bf v}_{i}{\partial P_{N}\over\partial {\bf r}_{i}}+{\bf F}_{i}{\partial P_{N}\over\partial {\bf v}_{i}}\biggr )=\sum_{i=1}^{N} {\partial\over\partial {\bf v}_{i}}\biggl\lbrack D  {\partial P_{N}\over\partial {\bf v}_{i}}+\xi P_{N}{\bf v}_{i}\biggr\rbrack,\nonumber\\
\end{eqnarray}
where ${\bf F}_{i}=-\nabla_{i}U({\bf r}_{1},...,{\bf r}_{N})$. We note
that the stationary solutions of this equation correspond to the
canonical equilibrium distribution (\ref{bgk16}) provided that
$\xi=D\beta$ (Einstein relation).  Integrating over ${\bf
r}_{2},{\bf v}_{2},...,{\bf r}_{N},{\bf v}_{N}$, we obtain
\begin{eqnarray}
\label{b11}
{\partial P_{1}\over\partial t}+{\bf v}_{1}{\partial P_{1}\over\partial {\bf r}_{1}}+\int \prod_{i=2}^{N}d^{D}{\bf r}_{i}d^{D}{\bf v}_{i} \ {\bf F}_{1}{\partial P_{N}\over\partial {\bf v}_{1}}= {\partial\over\partial {\bf v}_{1}}\biggl\lbrack D {\partial P_{1}\over\partial {\bf v}_{1}}+\xi P_{1}{\bf v}_{1}\biggr\rbrack.\nonumber\\
\end{eqnarray}
Now,
\begin{eqnarray}
\label{b12}
\int \prod_{i=2}^{N}d^{D}{\bf r}_{i}d^{D}{\bf v}_{i}\ {\bf F}_{1}{\partial P_{N}\over\partial {\bf v}_{1}}=-\int \sum_{j=2}^{N}\prod_{i=2}^{N}d^{D}{\bf r}_{i}d^{D}{\bf v}_{i} \ {\partial u\over\partial {\bf r}_{1}}({\bf r}_{1}-{\bf r}_{j}){\partial P_{N}\over\partial {\bf v}_{1}}\nonumber\\
=-(N-1)\int d^{D}{\bf r}_{2}d^{D}{\bf v}_{2} \ {\partial u\over\partial {\bf r}_{1}}({\bf r}_{1}-{\bf r}_{2})
{\partial P_{2}\over\partial {\bf v}_{1}}({\bf r}_{1},{\bf v}_{1},{\bf r}_{2},{\bf v}_{2},t).
\end{eqnarray}
Implementing a mean-field approximation which is valid in the
proper thermodynamic limit $N\gg 1$ with $\beta N u\sim 1$ \cite{bbgky}:
\begin{equation}
\label{b13}
P_{2}({\bf r}_{1},{\bf v}_{1},{\bf r}_{2},{\bf v}_{2},t)=P_{1}({\bf r}_{1},{\bf v}_{1},t)P_{1}({\bf r}_{2},{\bf v}_{2},t)+O(1/N),
\end{equation}
we find that
\begin{equation}
\label{b14}
{\partial P_{1}\over\partial t}+{\bf v}_{1}{\partial P_{1}\over\partial {\bf r}_{1}}+\langle {\bf F}\rangle_{1}{\partial P_{1}\over\partial {\bf v}_{1}}= {\partial\over\partial {\bf v}_{1}}\biggl\lbrack D {\partial P_{1}\over\partial {\bf v}_{1}}+\xi P_{1}{\bf v}_{1}\biggr\rbrack,
\end{equation}
where 
\begin{equation}
\label{b15}
\langle {\bf F}\rangle_{1}=-N\int d^{D}{\bf r}_{2}d^{D}{\bf v}_{2} \ {\partial u\over\partial {\bf r}_{1}}({\bf r}_{1}-{\bf r}_{2})P_{1}({\bf r}_{2},{\bf v}_{2},t).
\end{equation}
This can be rewritten
\begin{equation}
\label{b16}
{\partial f\over\partial t}+{\bf v}{\partial f\over\partial {\bf r}}+\langle{\bf F}\rangle {\partial f\over\partial {\bf v}}={\partial\over\partial {\bf v}} \biggl \lbrack D\biggl ({\partial f\over\partial {\bf v}}+\beta f{\bf v}\biggr )\biggr\rbrack,
\end{equation}
where 
\begin{equation}
\label{b17}
\langle{\bf F}\rangle=-\nabla\Phi=-\int d^{D}{\bf r}' \ \rho({\bf r}',t){\partial u\over\partial {\bf r}}({\bf r}-{\bf r}'),
\end{equation}
is the mean-field force created by the particles.

\end{document}